\documentstyle[epsf]{mn}
\begin{document}

\title[Clusters and MOND]{Clusters of galaxies with modified Newtonian 
dynamics (MOND)}

\author[R.H. Sanders] {R.H.~Sanders\\Kapteyn Astronomical Institute,
P.O.~Box 800,  9700 AV Groningen, The Netherlands}

 \date{received: ; accepted: }

\maketitle

\begin{abstract}

X-ray emitting clusters of galaxies are considered in the context of 
modified Newtonian dynamics (MOND).  I show that self-gravitating 
isothermal gas spheres are not good representations of rich clusters 
with respect to the radial gas density distribution as indicated
by the X-ray surface brightness.
Pure gas spheres with a density distribution described by a $\beta$ model,
as observed, also fail because, with MOND, these objects are far 
from isothermal and 
have a gas mass and luminosity much larger than observed for clusters of the
same mean temperature.  These problems may be resolved by adding an additional
dark mass component in the central regions;  a constant density 
sphere contained
within two core radii and having a total mass of one to two times
that in the gas.  With this additional component, the observed 
luminosity-temperature relation for clusters of galaxies is reproduced.
When the observed X-ray surface brightness distribution in actual clusters is 
modeled by such a 
two-component structure, the typical mass discrepancy is three to four times
smaller than with Newtonian dynamics.  Thus while MOND significantly reduces
the mass of the dark component in clusters it does not remove it completely.
I speculate on the nature of the dark component and argue that this is not
a fundamental problem for MOND.

\end{abstract}


\section {Introduction}

The modified Newtonian dynamics (MOND) is an empirically-based modification
of Newtonian gravity or inertia in the limit of low accelerations 
($< a_o \approx cH_o$) suggested
by Milgrom (1983) as an alternative to cosmic dark matter.  In addition to
explaining galaxy scaling relations (Tully-Fisher, Faber-Jackson, 
Fundamental Plane),
this simple algorithm allows one to accurately predict the shapes of spiral
galaxy rotation curves from the observed distribution of gaseous and stellar
matter.  MOND also accounts for the kinematics of small groups of galaxies
(Milgrom 1998) and of superclusters, as exemplified by the Perseus-Pisces
filament (Milgrom 1997) without the need for unseen mass. These 
well-documented phenomenological successes  
(Sanders \& McGaugh 2002 and references therein) challenge the cold dark 
matter (CDM) paradigm and provide
some support for the suggestion that the current theory of gravity and
inertia (General Relativity) may need revision in the limit of low
accelerations or field gradients.

However, problems do arise when one attempts to apply MOND 
to the large clusters of galaxies.  The and White (1988) first noted that,
to successfully account for the discrepancy between the observed mass
and the traditional virial mass in the Coma Cluster, the MOND acceleration
parameter, supposedly a universal constant, should be about a factor
of four larger than the value implied by galaxy rotation curves.  With MOND,
the dynamical mass of a pressure supported system at temperature T 
is $M\propto {T^2}/a_o$; therefore, the The and White result could also be
interpreted as an indication that the MOND dynamical mass is still larger
than the detectable mass in stars and gas.  

In astronomical tests involving an individual extragalactic object, such as the
Coma cluster, a contradiction is not necessarily a falsification. One can
always argue that the peculiar aspects of an object, such as deviations from
spherical symmetry or incomplete dynamical relaxation, exempt that particular 
case.  However, Gerbal et al. (1992), looking at a sample of eight X-ray
emitting clusters, noted that the problem is more general:  although MOND
reduced the Newtonian discrepancy by a factor of 10, there is still a need
for dark matter, particularly in the central regions.  Later, considering
a large sample of X-ray emitting clusters, I found that the mass predicted
by MOND remains, typically, a  factor of two or three times larger than the
total mass observed in the hot gas and in the stellar content of the galaxies
(Sanders 1998).  More recently,  Aguirre, Schaye \& Quataert (2001) pointed
out that MOND is inconsistent with  the observed temperature gradient in
inner regions of three clusters for which such data is available.  Again, the
problem can be remedied by additional  non-luminous mass, primarily in the
inner regions, of order two or three times the observed gas mass.  This
discrepancy is also evident from strong gravitational lensing in the central
regions of clusters-- the formation of multiple images of background
galaxies.  Here, MOND does not apply because accelerations are Newtonian, and 
the implied surface density greatly exceeds that of visible matter and 
hot gas (Sanders 1998).  
So, although MOND clearly reduces the  classical Newtonian mass discrepancy in
clusters of galaxies, there still  remains a missing mass problem,
particularly in the cores.  

Here I consider the issue of the remaining missing mass in clusters and 
whether or not this a fundamental problem for MOND.  First I calculate
the structure and X-ray surface brightness distribution of MOND isothermal
gas spheres.   Except for the very central regions, 
the structure of these objects is self-similar; the 
finite mass is primarily determined by the 
temperature and is very weakly dependent upon the central density.
Thus there exists a mass-temperature relation ($M\propto T^2$)
which is absolute; this implies a gas mass 
typically 5 to 10 times larger than that observed in X-ray emitting clusters
of the same temperature.
Moreover, for central electron densities in the range of
0.001 to 0.01, there is a well-defined X-ray luminosity-temperature relation 
which is less steep ($L\propto T^{1.5}$) and lies well 
above the observed luminosity-temperature relation for clusters 
($L\propto T^{2.5}$).  Looking at individual objects, the 
radial dependence of X-ray surface brightness does not 
reproduce that typically observed in X-ray emitting clusters-- observations
which are well-fit by the traditional ``$\beta$-model" (Sarazin 1988). 
The conclusion is that self-gravitating MOND isothermal gas spheres are 
not good representations of clusters of galaxies.

Gas spheres with a density distribution described by a 
$\beta$-model are not isothermal in the context of MOND. 
A central or boundary temperature must be specified to determine 
the run of temperature in such objects, but those models with
the lowest temperature gradients have large core radii and are again
over-massive and over-luminous with respect to observed clusters of the 
same temperature.
These problems may be solved by postulating the existence of a second 
rigid component in the mass distribution:  
a constant density core having
a radius about twice that of the $\beta$ model core and a central surface 
density comparable to $a_o/G$.  Here, by rigid I mean a component with a fixed
density distribution which does
not respond to the gravitational field of the hot gas or galaxies.   
The presence this additional component in the Coma
cluster is implied by the MOND hydrostatic gas equation for cumulative mass.
If this component is generally present in clusters, it contributes to the 
gravitational force in the inner regions and 
reduces the core radius at a given temperature.  In this way the observed 
luminosity-temperature and mass-temperature relations for clusters may be 
reproduced.  

About 40 individual clusters have been considered in terms of such a 
two-component model; i.e.,
the observed surface-brightness distributions and mean temperatures 
are fit by specifying the density of the non-luminous component which
is assumed to extend to two gas core radii.
The total mass of this additional component varies between a few times
$10^{12}$ and $10^{14}$ M$_\odot$ and the implied mass-to-light ratio is
typically 50 in solar units.  Therefore, the required rigid component
is not a standard stellar population; as noted previously, MOND
requires dark, or heretofore undetected, matter in the central regions of 
rich clusters.  Although the total discrepancy between dark and detectable
mass is reduced by, on average, a factor of four over that implied by 
Newtonian dynamics, it is clear that there remains a dark matter problem
for MOND.  I discuss the issue of whether or not this is a contradiction.
I speculate on the possible nature of this non-luminous component 
and argue that neutrinos of finite mass are a possible candidate.

\section {The structure and properties of MOND gas spheres}

\subsection {MOND isothermal spheres}

The structure of isotropic isothermal spheres may be determined by solving
the equation of hydrostatic equilibrium:

$${dp\over{dr}} = -\rho g \eqno(1a)$$
with the pressure $p$ given by 
$$p = \rho{\sigma_r}^2\eqno(1b)$$
where $\sigma_r$ is the one-dimensional velocity dispersion (constant for
an isothermal sphere), $\rho$
is the density.  The gravitational acceleration, $g$, in the
context of MOND, is given by
$$g\mu(g/a_o) = g_n. \eqno(2a)$$
where $g_n$ is the usual Newtonian gravitational acceleration,
$$g_n = {{GM(r)}\over r^2}, \eqno(2b)$$ 
$a_o$ is the critical acceleration below which gravity
deviates from Newtonian (found to be about $10^{-8}$ cm/s$^{-2}$ from fits
to galaxy rotation curves), and $\mu(x)$ is a function which interpolates
between the Newtonian regime ($\mu(x)=1$ when $x>>1$) and the MOND
regime ($\mu(x)=x$ when $x<<1$).  A function having this asymptotic form, 
$$\mu(x) = {x(1+x^2)^{-{1\over 2}}}, \eqno(2c)$$ works well for galaxy rotation
curves and is also used here.

Milgrom (1984) demonstrated that MOND isothermal spheres have a finite mass
and a density which falls of as $r^{-\alpha}$ where $\alpha \approx 4$
at large radii.  In the outer regions, where $M(r)=M$ = constant, it follows
immediately from eqs.\ 1 and 2 that
$${\sigma_r}^4 = {\alpha^{-2}}GMa_o\eqno(3)$$
which means that the mass is uniquely determined by the specified velocity
dispersion or temperature.  For an isothermal gas sphere this relation becomes 
$$M \approx {16\over{Ga_o}}{\Bigl({{kT}\over{fm_p}}\Bigr)}^2 = 
(2.9\times 10^{13}) {T_{keV}}^2 \,M_\odot\eqno(4)$$
where $f$ is the mean atomic weight ($\approx 0.62$ for an fully ionized gas
with solar abundances) and $m_p$ is the proton mass.  

This would be, in effect, the extension of the Faber-Jackson relation to
clusters of galaxies (Sanders 1994).  However, the observational definition
of such a relation is ambiguous because the total gas mass defined by the
$\beta$-model is typically divergent.  If one considers the mass inside
a fixed radius (Sanders 1994), or within a radius where the density falls
to some fixed value (Mohr, Mathiesen \& Evrard 1999), a relationship of 
this form (eq.\ 4) is observed for clusters.  However, the mass at a given
temperature is typically a factor of 5 to 10 below that implied by
eq. 4.  This is shown in Fig.\ 1 where the MOND mass-temperature relation 
for isothermal gas spheres (dashed line) 
is compared to the observed gas mass-temperature
relation for 42 clusters listed in Table 1.  Here the observed gas mass is
given by the $\beta$-model (eq.\ 7 below)
extrapolated to a radius where the gas density has
fallen to about $10^{-28}$ g/cm$^3$, or 250 times the mean cosmological
density of baryonic matter.

\begin{figure}
\epsfxsize=80mm
\epsfbox{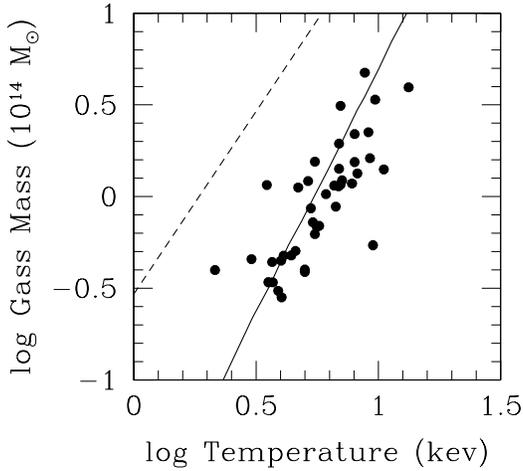}
\caption{The gas mass-temperature relation for clusters of galaxies.  The points
are measured temperatures and inferred gas masses for the 42 clusters listed 
in Table 1.  The gas mass is that obtained by extrapolating the 
$\beta$-model to a radius where the gas density falls to $10^{-28}$ g/cm$^3$.
The dashed line is the mass-temperature relation for MOND 
isothermal spheres (eq.\ 4) and the solid curve is the gas mass-temperature
relation for the two component MOND $\beta$ models discussed in Section 3.}
\hfil
\end{figure}

The characteristic scale of MOND isothermal gas spheres is
$$r_m\approx {\sigma_r}^2/a_o\eqno(5)$$ (Sanders \& McGaugh 2002).  Given that the
X-ray luminosity due to free-free emission is 
$L\propto {n_e}^2T^{1\over 2}{r_m}^3$;
then, eqs. 4 and 5 above would imply an X-ray luminosity-temperature relation
of the form $L\propto T^{1.5}$.  The observed luminosity-temperature 
relation (e.g., Ikebe et al. 2002) is significantly steeper,
$L\propto T^{2.5}$, than that of MOND isothermal spheres.

To pursue this in more detail, I calculated the X-ray luminosity 
for MOND isothermal spheres with electron densities and temperatures 
similar to those of the X-ray emitting clusters of
galaxies.  I assume a central electron density of 0.006 cm$^{-3}$ as
being typical of clusters, and, for a given temperature, numerically integrate 
eqs. 1 and 2 from the center outward.  The optically thin
free-free radiation for the entire sphere is then calculated from the
run of electron density.  The resulting luminosity-temperature relation
is shown by the dashed curve in Fig.\ 2.  compared to the observations
by Ikebe et al. (2002) scaled to h=0.7.  In both cases, this is the radiation
emitted in the band 0.2 to 2.4 keV where the X-ray flux is typically measured
by satellites such as ASCA and ROSAT. Not only is the theoretical dependence
shallower than observed, the predicted luminosities are an order of
magnitude larger than actual clusters at the same temperature. 

\begin{figure}
\epsfxsize=80mm
\epsfbox{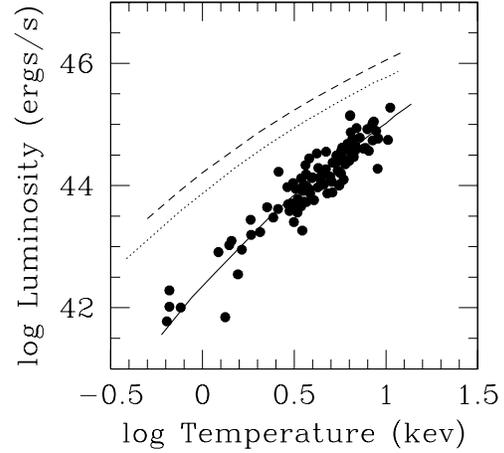}
\caption{The X-ray luminosity-temperature relation for clusters of galaxies.
The points are the clusters tabulated by Ikebe et al. scaled to h=0.7.  The 
dashed curve is the relation for MOND isothermal spheres and the dotted curve
is the relation for the near-isothermal MOND $\beta$ models.  The solid curve
is the relation for the MOND two-component models discussed in Section 3.}
\vfill
\end{figure}

On the basis of these scaling relationships, it would seem that MOND
isothermal spheres are not good representations of clusters of galaxies.
This conclusion is reinforced when we consider the surface brightness 
distribution of a MOND isothermal gas sphere.  Although the mass of an isotropic,
isothermal sphere is effectively determined by the temperature, the detailed
structure depends
upon the central density;  for a single temperature there is a family
of solutions bounded by a limiting solution with a 1/r density cusp at the
center (Milgrom 1984).  The limiting MOND solution resembles the Hernquist
model (Hernquist 1990) with the 1/r cusp steepening to a 1/r$^4$ 
beyond a radius $r_m$.
Lower density spheres are characterized by a constant density core with
a 1/r$^4$ gradient at large r;
the central gas densities observed in clusters of galaxies would place
these objects in this category of the sub-limiting solutions.  The X-ray surface
brightness distribution resulting from such a MOND isothermal sphere is
shown in Fig.\ 3 compared to that observed for the Coma cluster (Briel, Henry
\& B\"ohringer 1992).  
These observations are well fit by the traditional $\beta$-model
(Cavaliere \& Fusco-Femiano 1976, Sarazin 1988),
$$I(r) = I_o{\Bigl[1 + \Bigl({r\over{r_c}}\Bigr)^2\Bigr]}^{-3\beta
+1/2}\eqno(6)$$ shown by the dashed curve in Fig.\ 3.  Here we see that the
MOND isothermal gas sphere provides a poor representation of the actual surface
brightness distribution.  Moreover, the temperature of this best fitting
MOND isothermal sphere is about 2.5 keV whereas the actual temperature of
the Coma cluster is in excess of 8 keV.

\begin{figure}
\epsfxsize=80mm
\epsfbox{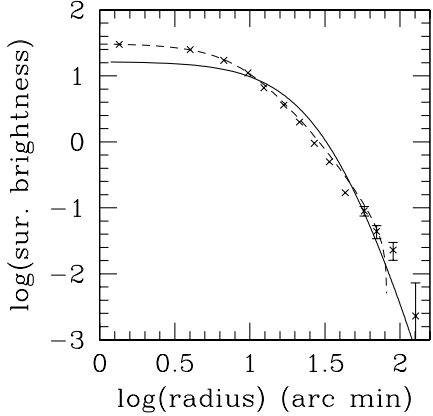}
\caption{The radial distribution of X-ray surface brightness for a MOND
isothermal sphere (T=2.5 keV) compared to observations of the Coma cluster
(T=8.6 keV).  The dashed curve is the $\beta$ model fit (Reiprich 2000) to
the observations.}
\vfill
\end{figure}

\subsection {MOND $\beta$-models}

The radial dependence of electron number density which produces the X-ray 
intensity distribution described by eq.\ 6 is 
$$ n_e = n_o{\Bigl[1+\Bigl({r\over{r_c}}\Bigr)^2\Bigr]}^{-1.5\beta}\eqno(7)$$
(Cavaliere \& Fusco-Femiano 1976).
Rather than assuming a constant temperature and solving eq.\ 1 for the
density distribution, one may alternatively 
take eq.\ 7 as the density distribution 
and solve for the radial dependence of the temperature.  
It is necessary to specify $\beta$ 
(typically 0.6 to 0.7 for clusters), a central electron
density, $n_o$ (again taken to be  0.006 cm$^{-3}$), a core 
radius, $r_c$ (ranging from 50 to 300 kpc for clusters), 
and a central gas temperature, $T_o$.  Then eqs.\ 1 and 2 may be solved 
for the run of temperature to an outer boundary, usually taken to be where
the  gas density falls to some fixed multiple of the mean cosmological
density.   For a given central temperature, the run of temperature is
completely determined by the core radius.  There is one specific value of
$r_c$  which minimizes the temperature gradients.  For smaller core radii, 
the temperature rapidly increases toward the boundary (the models are very 
far from isothermal);  for larger core radii, the temperature decreases to 
zero before the boundary is reached.  Because clusters of galaxies are
observed to be
near isothermal, these MOND $\beta$-models with the smallest temperature
variations are taken to be appropriate for clusters.

\begin{figure}
\epsfxsize=80mm
\epsfbox{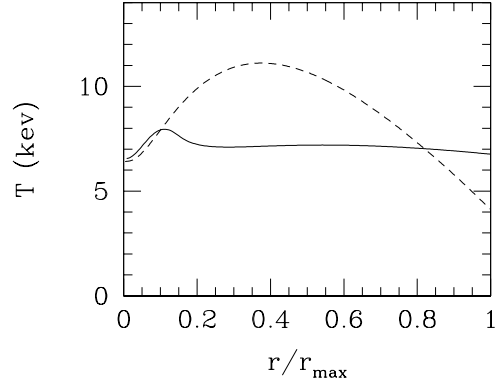}
\caption{The projected emission-weighted temperature of a MOND $\beta$ model 
as a function of projected radius (dashed curve).  The radius is given
in terms of $r_{max}$ where the gas density has fallen to $10^{-28}$
g/cm$^3$.  This is the most nearly
isothermal $\beta$ model.  The solid curve shows the projected 
emission-weighted temperature for the most nearly isothermal 
two component model described in Section 3.  }
\vfill
\end{figure}

The emission-weighted temperature of this most nearly-isothermal model
as a function of {\it projected} radius is shown by the dashed curve
in Fig.\ 4 for $n_o = 0.006$ cm$^{-3}$, $\beta = 0.62$, and $T_o=6.5$ keV.
The required core radius is 436 kpc.  It is evident from Fig.\ 4 that 
the model still deviates significantly from pure isothermal, with a roughly a
factor of two variation around the central temperature.  Moreover, compared
to actual clusters, the implied core radius for this temperature is too large 
by a factor of two.  

For these near isothermal $\beta$ models, I determined the mean emission
weighted temperature, the total gas mass within the cut-off radius
and the X-ray luminosity within the 0.1 to 2.4 keV band.
The resulting gas mass-temperature relation is almost identical to
that of MOND isothermal spheres (the dashed curve in Fig.\ 1) 
and thus again much larger than the observationally inferred gas mass 
of X-ray emitting clusters 

The luminosity-temperature relation for these MOND $\beta$ models is
shown in Fig.\ 2 by the dotted curve. Unsurprisingly, this nearly coincides 
with the calculated relation for MOND isothermal spheres 
and is clearly an equally
poor description of reality.  The basic problem is that, for both MOND
isothermal spheres and MOND $\beta$ models, the core radii are too large.  It
is evident that some ingredient is missing from these models;  that an
additional mass component must be added to decrease the gas core radius at a
given temperature.

\section {Resolution of the problem:  two component models}

For a cluster with an observed density and temperature distribution, eqs.\ 1
and 2 directly yield M(r), the interior dynamical mass as a function of
radius.  In the Newtonian regime this is given simply by  
$$M_N(r) = {{r}\over G} {{kT}\over fm_p}
\Bigl[{{d\,ln(\rho)}\over{d\,ln(r)}} +
{{d\,ln(T)}\over{d\,ln(r)}}\Bigr]_r.\eqno(8)$$ With MOND, taking $\mu(x)$ to be
given by eq.\ 2c, the dynamical mass is
 $$M_m = {{M_N}\over{\sqrt{1+(a_o/a)^2}}} 
\eqno(9)$$ where a is the ``observed" acceleration 
$$ a = {1\over\rho}{{dp}\over{dr}}=
{{kT}\over{fm_pr}}\Bigl[{{d\,ln(\rho)}\over{d\,ln(r)}}+ {{d\,ln(T)}\over 
{d\,ln(r)}}\Bigr]_r.\eqno(10)$$ 
Obviously, from eq.\ 9 in the limit of large accelerations ($a>>a_o$) the 
MOND dynamical mass is equivalent to the Newtonian dynamical mass.

\begin{figure}
\epsfxsize=80mm
\epsfbox{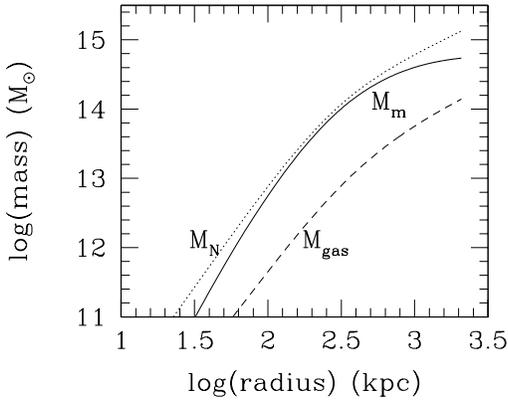}
\caption{The accumulated mass distributions in the Coma cluster.
The dotted curve is the enclosed Newtonian mass as a function of radius;
the solid curve is the enclosed MOND mass; the dashed curve is the gas
mass inferred from the X-ray observations}
\vfill
\end{figure}

We may apply these relations to the Coma cluster which has a density distribution
well-fit by a $\beta$-model with $\beta$ = 0.71, $r_c$ = 276 kpc, $n_o$ =
.0036 (Reiprich 2001), an observed radial temperature profile (Arnaud et al.
2001) and average emission weighted temperature of 8.6 keV.
The results are shown in Fig.\ 5 which is the cumulative Newtonian
dynamical mass (dotted line), the MOND dynamical mass (solid line), and
the gas mass (dashed line).  Here it is evident that while the Newtonian
dynamical mass continues to increase at radius of 1 Mpc, the MOND mass has
essentially converged.  However, the total MOND mass is still a factor of
four times larger than the mass in gas alone.  This discrepancy cannot be
accounted for by the stellar content of the galaxies which, assuming a 
mass-to-light ratio of seven,
amounts only to about $10^{13}$ M$_\odot$ within 1 Mpc (The \& White 1988). 
This is, in fact, the discrepancy pointed out by The and White-- a discrepancy
which can be resolved by increasing $a_o$ by a factor of 3 or 4 over the value
required for galaxy rotation curves, or by admitting the presence of
non-luminous mass which MOND does not remove.

The density of this non-luminous component is roughly constant and contained 
within two gas core radii.  In other words the 
missing mass is essentially present in the inner regions of the cluster
as implied in the work of Aguirre, Shaye, and Quataert (2001).
This suggests that, with MOND, clusters might be described by a two component
model:  a gas component with a density distribution given by the 
$\beta$-model and a dark central component of constant
density and a radius of about two times the core radius of the
the gas distribution. 

In Coma, the central surface density of the dark component is about 
$\Sigma_d = 240\,\,M_\odot$/pc$^2$, which is comparable to the MOND
surface density of $a_o/G \approx 700\,\,M_\odot$ pc$^2$.  
This is the characteristic
central surface density of MOND self-gravitating isothermal systems 
(Milgrom 1984).  Therefore, to determine scaling relations, I assume that  
the non-luminous component is a rigid sphere having a central 
surface surface density equal to that in the Coma cluster   
and radius twice that of the gas core radius, as in Coma.
Then the constant density of the second component is
$$\rho_d = {\Sigma_d\over {2r_c}}\eqno(11)$$
and the total dark mass is
$$M_d = {{16\pi}\over{3}}\Sigma_d{r_c}^2\eqno(12)$$
Evidently, with this assumption, larger clusters have a more massive dark
component, but because the gas mass scales as ${r_c}^3$, the dark to gas 
mass decreases with increasing core radius or temperature.

The procedure followed is identical to that of the pure MOND $\beta$ models
described above:  I assume a gas distribution described by eq.\ 
with $\beta = 0.62$, $n_o = 0.006$.  Then for a given central gas temperature,
I determine the core radius of the model for which the temperature gradient
is minimized.  The new aspect is the second component
which makes its presence felt by contributing to the
total cumulative mass (M(r) in eq.\ 2) and hence the total gravitational
force.  This has the effect of decreasing
the core radius at a given temperature compared to the single component
MOND $\beta$ models.  The
emission-weighted temperature of as a function of projected radius is shown by
the solid line in Fig.\ 4 again for a model with a  central temperature of 6.5
keV.  This is the most-nearly isothermal model, and we see that the
temperature gradients are much smaller than in the the single component
$\beta$-model.  The total variation about the central temperature is less than
40\%.  In other words, {\it isothermal $\beta$ models require, 
in the context of MOND, this second central component with roughly 
constant density}.

The gas-mass-temperature relation of such models is shown by the solid line 
in Fig.\ 1 which is evidently consistent with the observations.    
The luminosity-temperature relation
for these two component cluster models is shown by the solid line in Fig.\ 2.
These models provide a reasonable description of
the observed relation.  This is due to the fact that, in the low temperature
systems, a relatively larger fraction of the mass is not in gaseous form.
It is also not in the form of luminous material in galaxies as the implied
mass-to-light ratios would be too large.  Agreement of MOND with the cluster
scaling relations is achieved at the expense of adding unseen matter.

\section {Modeling individual clusters}

Individual clusters may be described by such two component models.  
Given the parameters of a $\beta$ model fit to the X-ray emission from an
specific cluster (i.e., $\beta$, $n_o$, and $r_c$), and a characteristic
temperature for the entire cluster, I determine, via eqs. 1 and 2, the central
temperature, $T_o$, and the density (or surface density, related by eq.\ 11) 
of the dark component
which yields the observed emission weighted temperature and the observed
core radius, $r_c$ of the $\beta$ model.  In all cases the dark component
is assumed to extend to 2$r_c$.  Uniqueness is ensured by requiring
that the temperature gradients be minimized-- i.e., these are again the
most nearly isothermal models.  The fitting parameter is the density
of the central dark component which yields the total dark mass via eq.\ 12
above.

Table 1 lists the clusters which have been modeled in this way along with
the the parameters of the $\beta$-model fit and the mean temperature, all 
from the compilation of Reiprich (2001) with cluster properties scaled
to h=0.7.  The the required central surface
density of the dark component, $\Sigma_d$, is given along
with the total gas mass (out to the cutoff radius) and the total mass
of the dark component. The enclosed Newtonian dynamical mass is also given. 

\begin{table*}
\begin{minipage}{80mm}
\caption{ Two component model fits to observed clusters\label{t1}}
\begin{tabular}{|c|c|c|c|c|c|c|c|c|c|}
${\rm Cluster }$ &${\rm T }$ & ${\beta}$ & ${\rm r_c}$ & ${\rm n_o}$ &
   ${\Sigma_d}$
   &$ {\rm {\dot M}_c}$ &  ${\rm M_g}$ & ${\rm M_d}$ & $ {\rm M_N}$ \\
   $  $ & $ {\rm keV} $ &   $ $ &
   ${\rm kpc} $ & ${\rm 10^{-3} cm^{-3}}$ & $ {\rm 100 M_\odot/{pc}^2} $ &
   $ {\rm M_\odot/year} $ & $ {\rm 10^{14} M_\odot} $ &
   ${\rm 10^{14} M_\odot}$ & ${\rm 10^{14} M_\odot} $  \\
 (1) & (2) & (3) & (4) & (5) & (6) & (7) & (8) & (9) & (10) \\

   A85 &  6.9 & 0.532 & 58.1 & 20.4 & 12.5 & 198 & 1.94 & 0.704 & 10.9 \\
  A119 &  5.6 & 0.675 & 359. & 1.25 & 0.8  & 0   & 0.679 & 1.73 &  7.27 \\
  A262 &  2.15 & 0.443 & 30. & 9.57 & 4.5  & 27  & .397  & 0.0678 & 2.83 \\
  A399 &  7.0  & 0.713 & 320.&  2.54 & 1.62 & 0  &  1.17 & 2.77 &  10.7 \\
  A401 &   8.0 & 0.613 & 175.&  6.72 & 3.9  & 42 &  2.19 & 1.99 &  13.8 \\
  A576 &   4.02& 0.825 & 283.&  1.88 & 0.99 & 69 &  0.275& 1.35 &   4.17 \\
  A754 &   9.5 & 0.698 & 171.&  5.21 &  6.1 & 216&  0.589& 2.94 &   8.09 \\
 A1367 &   3.55& 0.695 & 273.&  1.51&    .6&  0  & 0.35 &  .738 &  3.82  \\
 A1644 &   4.7 & 0.579  & 214.&  2.60&  1.03&  12 & 1.12 & 0.789 &  7.64  \\
 A1651 &   6.1 &  0.693 & 129. &  9.26&   4.45&  138 &   1.05&  1.22&  7.76 \\
 A1650 &   6.7 &  0.704 &  201.&  5.08&  2.85 & 122  & 0.893 &  1.92& 8.10  \\
 A1656 &   8.21&  0.705 &  275.&  3.64&  2.6  &   0 &  1.36 & 3.27 &  12.2 \\
 A1689 &   9.23&  0.690 & 116. & 20.2 &  9.15 &  484 & 1.60 & 2.08 &  11.0 \\
 A1736 &   3.5 &  0.542 &  267.&  1.51&  0.23 &  0   & 1.16 & 0.275&   6.87 \\
 A1795 &   7.8 &  0.596 &  56.5&  30.3&   18. &  381 & 1.18 & 0.96 &   7.97 \\
 A1914 &  10.53&  0.751 &  167.& 13.2 &    7.5&      & 1.43 & 3.48 &  12.1 \\
 A2029 &   9.1 &  0.582 &  59.3& 34.2 &   20.0&  556 & 2.21 & 1.21 &  12.5 \\
 A2052 &   3.03&  0.526 &  26.4& 31.7 &  13.5 &  94  & 0.459& 0.154&   3.24 \\
 A2063 &   3.68&  0.561 &  78.6&  7.42&    4.0&   37 & 0.447& 0.407&   3.74 \\
 A2065 &   5.5 & 1.162 &  496.&  2.36&  1.38 &  0   & 0.619& 5.69 &  10.5 \\
 A2142 &   9.7 &  0.591 & 110. & 16.3 &    9.5&  286 & 3.35 & 1.95 &  18.0 \\
 A2163 &  13.29&  0.796 &  371.&  6.17&    3.7&    0 & 3.93 & 8.54 &  27.3 \\
 A2199 &   4.1 &  0.655 &  99.3&  9.84&    3.7&   94 & 0.475& 0.612&   4.08 \\
 A2244 &   7.1 &  0.607 &  90. & 14.2 &    8.8&  244 & 1.19 & 1.23 &   8.51 \\
 A2255 &   6.87&  0.797 & 424. &  1.95&    1.2&    0 & 1.12 & 3.63 &  11.5 \\
 A2256 &   6.6 &  0.914 & 419.&  3.1 &    1.4&    0 &  1.19&  4.08&   11.7\\
 A2319 &   8.8 &  0.591 & 204. &  6.96&    2.9&   20 &  4.80&  2.02&   23.2\\
 A2597 &   4.4 &  0.633 & 41.4 & 43.2 &   13. &  271 & 0.476& 0.374&   3.59 \\
 A2634 &   3.7 &  0.640 &  261.&  1.27&   .566&  0   & 0.339& 0.648&   3.69 \\
 A3112 &   5.3 &  0.576 & 43.6 & 33.1 &   15. &  376 & 0.869& 0.471&   5.75 \\
 A3266 &   8.0 &  0.796 &  403.&  2.74&   1.58&   4  &  1.57&  4.21&   14.0\\
 A3376 &   4.0 & 1.054 &  539.&  1.21&    0.5&   53 &  0.444& 2.44&    6.50\\
 A3391 &   5.4 &  0.579 &  167.&  3.04&   2.24&   0  &  0.706& 1.06&    6.32\\
 A3395n&   5.0 &  0.981 &  478.&  1.21&   0.89&      &  0.384& 5.02&    7.36\\
 A3395s&   5.0 &  0.964 &  431.&  1.51&   .99 &      & 0.405 & 3.07 &   7.14\\
 A3530 &   3.89&  0.773 &  301.&  1.51&     8.0 &      &  0.307& 1.21 &   4.27\\
 A3532 &   4.58&  0.653 &  201.&  2.74&    1.5&   0  &  0.507& 1.01 &   5.09\\
 A3558 &   5.5 &  0.580 &  160.&  5.17&    2.1&  40  &  1.56 & 0.894&   9.60\\
 A3562 &   5.16&  0.472 &  70.7&  6.69&    5.2&   37 &   1.22& 0.432&   7.64\\
 A3571 &   6.9 &  0.613 &  129.&  8.75&    5.0&   81 &   1.40&  1.41&   9.70\\
 A3667 &   7.0 &  0.541 &  199.&  4.01&    2.0&    0 &   3.13&  1.32&  16.7\\
 A3921 &   5.73&  0.762 &  237.&  4.06&    2.2&      &   0.697& 2.07&   7.30\\
\end{tabular}

\medskip
{All $\beta$-model parameters are from the compilation of Reiprich (2001). 
In columns 8 and 10 the masses are those enclosed within a radius
(typically 1 to 3 Mpc) where the gas density has fallen to 250 times its
mean cosmic value (assuming $\Omega_b = 0.04$).  In column 7 a blank entry
means that there is no estimate for the cooling mass inflow rate.}
\end{minipage}
\end{table*}

This sample was chosen to include a number of objects with significant
inferred cooling flows (such as Abel 1689 and 2029).  These clusters are
characterized by relatively small core radii ($r_c<100$ kpc) and large central 
electron densities ($n_o>0.01$ cm$^{-3}$).  The sample also includes clusters
at the other extreme-- large core radii ($r_c > 250$ kpc) and low central
electron densities ($n_o < 0.005$ cm$^{-3}$) with no inferred cooling flow
(such as Abel 119 and 2256).

In general, MOND reduces the classical Newtonian discrepancy
for clusters of galaxies.  For this sample, with MOND the ratio of the total 
dark mass to the gas mass is 1.60 $\pm$ 1.7.  
The Newtonian dark mass-to-gas mass ratio is 7.14 $\pm$ 2.7.  While this is
a significant reduction, it is also clear that MOND does not fully resolve the
mass discrepancy in clusters.  Moreover, it should be noted that the
ratio of masses of the dark-to-hot gas components in the central
two core radii (where the additional component is required) can be as
large as 10, as was also pointed out by Aguirre, Schaye \& Quataert (2001).  
This rigid component cannot
be stars of a normal population because the mass-to-light ratio within two
core radii would
still be, on average, in excess of about 50.  The principle question is
whether or not this dark component is a fundamental problem or can be
accommodated in the context of MOND.

\begin{figure}
\epsfxsize=80mm
\epsfbox{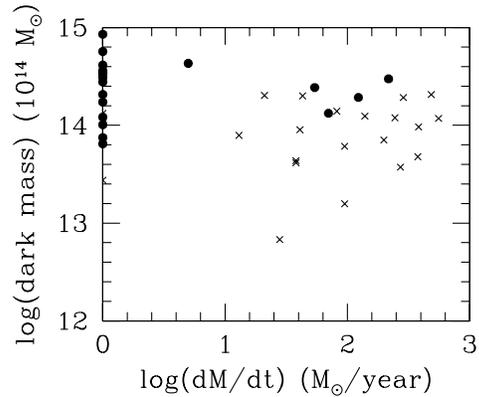}
\caption{A log-log plot of the fitted dark mass ($10^{14} M_\odot$) vs.
the cooling mass inflow 
rate inferred from central gas densities and temperatures
(White, Jones \& Forman 1997).  The crosses are those clusters with 
a small mass discrepancy ($M_d/M_g<1.5$), and the solid points are the
objects with large discrepancies ($M_d/M_g>1.5$).} 
\vfill
\end{figure}

It could be that a high M/L population of low mass stars or sub-stellar
objects is deposited in the central regions of clusters as a result of
cooling flows.
Fig.\ 6 is a plot of the total mass in the dark component vs. the cooling
rate as estimated by White, Jones \& Forman (1997).  The solid points are those
clusters with a large discrepancy:  $M_d/M_g>1.5$.
The crosses are clusters with a smaller discrepancy: $M_d/M_g<1.5$.
We see that there is no obvious correlation between the total mass of the
dark component and the cooling rate-- especially for those clusters with
the largest discrepancy.  On the other hand, in Fig.\ 7 we see a plot
of the surface density of the dark component vs. the mass deposition rate.
Here there does appear to be a correlation.  This is because those clusters
with the highest central dark matter densities are not the clusters with 
the largest mass discrepancies.

\begin{figure}
\epsfxsize=80mm
\epsfbox{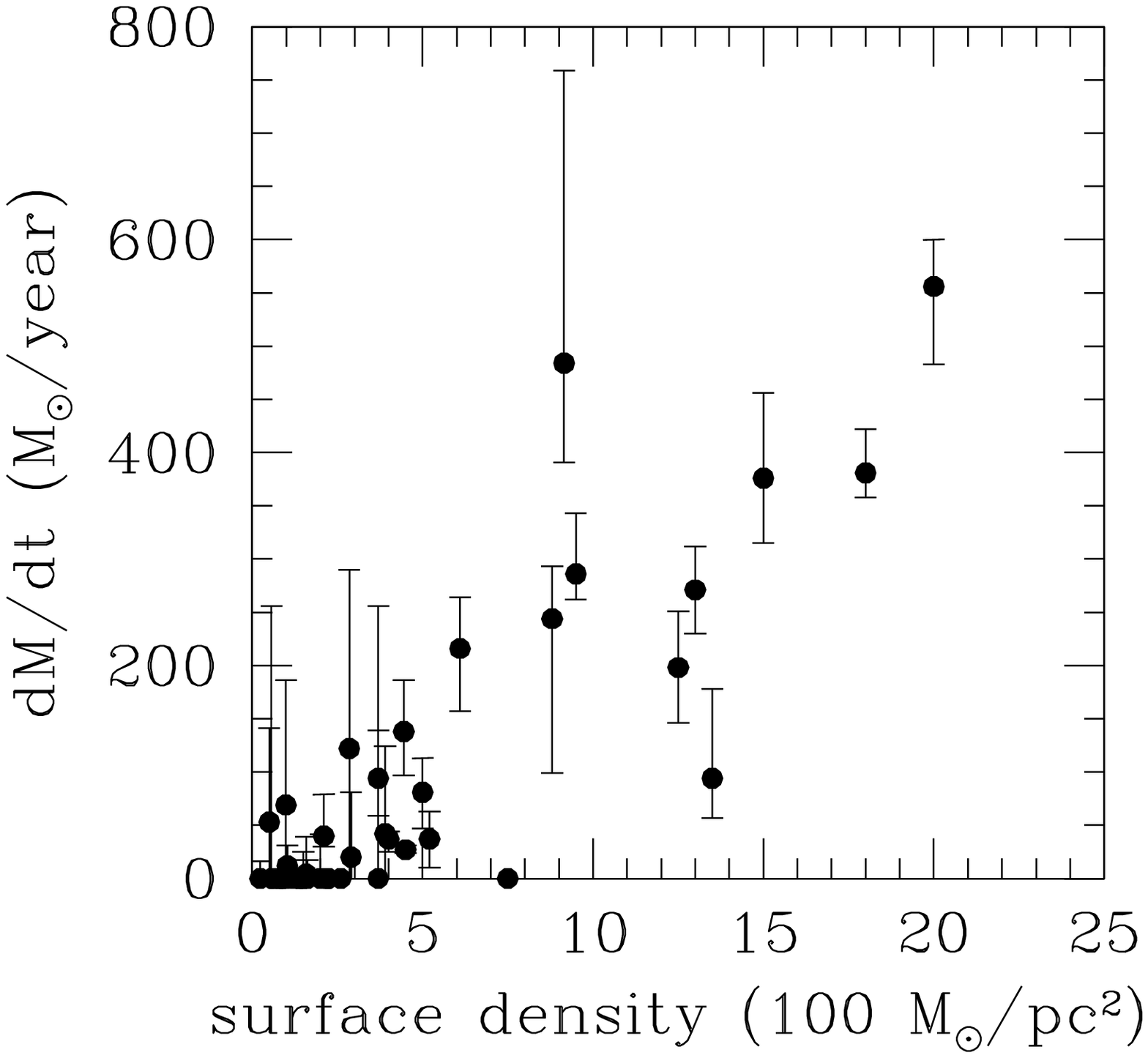}
\caption{The inferred cooling mass inflow rate in the sample clusters
(as in Fig.\ 6) vs. the central surface
density of the dark component in units of 100 M$_\odot$/pc$^2$.}
\vfill
\end{figure}

It is unclear if this apparent correlation between surface density and
mass deposition rate is significant.  The mass deposition is not actually
observed but calculated from the central gas density.  Those clusters 
with large inferred cooling flows are clusters with high central gas densities
and small core radii.  But it precisely these clusters which require a
large surface density of dark matter to produce the small core radius.
While it may be the case that cooling flows contribute to
the dark component of those clusters with the largest central density of
dark matter, it is also evident from Fig.\ 6 that this cannot be the explanation
for the discrepancy in clusters with the largest dark mass problem.  These
tend to be the clusters with low central gas densities and low inferred
dark matter densities-- but large core radii.  In the next section, I consider
another possibility:  particle dark matter in the form of massive neutrinos.

\section{Neutrinos as cluster dark matter}

The most well-motivated form of 
particle dark matter consists of neutrinos with finite mass.
Aspects of the observed fluxes of 
atmospheric and solar neutrinos provide strong evidence for neutrino oscillations
and hence non-zero neutrino masses (Gonzales-Garcia \& Nir 2002).
The fact that the number density of
neutrinos produced in the early universe is comparable to that of photons
then implies that there is a universal dark matter sea of neutrinos. 
The contribution to cosmological mass density would be
$\Omega_\nu h^2 = \sum{m_{\nu_i}}/94$ ev where the sum is over 
neutrino types. 

The neutrino oscillation experiments do not provide information on the
actual masses of neutrino species but on the square of the mass differences.
These are small, such that, the largest mass difference, suggested
by the atmospheric oscillations, is $\Delta m\approx
0.05$ ev.  If $m_\nu \approx \Delta m$ then the three active neutrino types
would have no significant cosmological mass density ($\approx 10^{-3}$)
and could not contribute to the mass budget of any bound astronomical object.  
But another possibility is that $m_\nu >>\Delta m$ and that the masses
of all three active types are nearly equal.  In this case an upper limit
to the masses is provided by an experimental limit on the mass of the
electron neutrino, i.e. 2.2 ev at 90\% c.l. (Groom et al. 2000)  
If it were the case that
the electron neutrino mass were near 2 ev, then neutrinos would constitute
a significant fraction of the cosmic density ($\Omega_\nu \approx 0.13$ 
for h=0.7).

However, neutrinos of this mass could not contribute to the mass budget
of galaxies.  This follows from a classic argument by Tremaine \& Gunn
(1979) based upon conservation of the phase space density of the
neutrino fluid.  Relativistic neutrinos are created with a maximum phase
space density of $(2\pi\hbar)^{-3}$ per type including anti-neutrinos
(this is a factor of two less than the 
absolute limit implied by quantum mechanical degeneracy).  In subsequent
evolution of the neutrino fluid involving gravitational instability and
collapse, the final phase space density cannot exceed this value.
This provides a relation between the final density of neutrino dark matter
and the velocity dispersion of the system; with three types
$$\rho_\mu \le (4.8\times 10^{-27}) \Bigl({{m_\nu}\over{2\,ev}}\Bigr)^4
  ({T_{keV}})^{3\over2} {\rm g/cm^3} \eqno(13)$$
Equivalently, for virialized
systems, this may be written as a relation between the effective core radius 
of a dark halo and its velocity dispersion; this is, roughly, 
$$r_d \ge 0.5\Bigl({{2\,ev}\over{m_\nu}}\Bigr)^2\Bigl({{1000\, {\rm km/s}}
\over{\sigma_r}}
 \Bigr)^{1\over 2}\, Mpc \approx 0.7 \Bigl({{2\,ev}\over{m_\nu}}
 \Bigr)^2({T_{keV}})^{-{1\over 4}}\, {\rm Mpc}\eqno(14)$$
Clearly for objects with the required velocity dispersion of galaxy halos 
($25\,{\rm km/s}<\sigma_r<200\, {\rm km/s}$), neutrinos in the mass range
of one to two ev 
could not possible cluster on sub-Mpc scales.  However, it obviously would
be possible for such neutrinos to contribute to the mass budget of large 
clusters.

\begin{figure}
\epsfxsize=80mm
\epsfbox{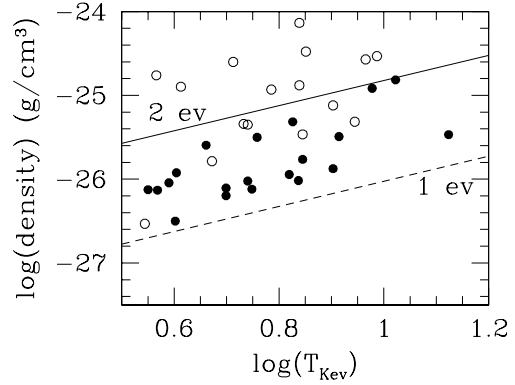}
\caption{A log-log plot of the fitted central density of the dark component
in the sample clusters vs. the temperature.  The solid line is the relation
between the maximum neutrino density and the temperature imposed by phase
space constraints for three neutrino types all with mass 2 ev (eq.\ 13).
The dashed line is the same when the neutrino mass is taken to be 1 ev.
The solid points are those clusters with a large discrepancy ($M_d/M_g>1.5$).}
\vfill
\end{figure}

Fig.\ 8 shows the the fitted matter density of the dark cores of the clusters
of galaxies listed in Table 1 vs. the gas temperature of the clusters.
The solid points are the clusters with significant mass discrepancies 
($M_d/M_g>1.5$) and the open points the clusters with lower discrepancies
($M_d/M_g<1.5$).  The solid and dashed lines show the relation between 
maximum neutrino density and the temperature implied by eq.\ 13, for 
neutrinos of 2 ev and 1 ev respectively.  For neutrinos of 2 ev,  
this relation appears to form an upper envelope for the clusters with the 
largest discrepancies.  However, due to the extreme sensitivity of this density
limit on neutrino mass (eq.\ 13), neutrinos with mass as low as 1 ev could not
comprise the dark component.

\begin{figure}
\epsfxsize=80mm
\epsfbox{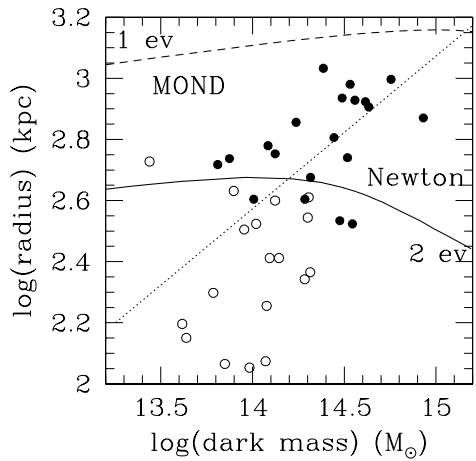}
\caption{The solid curve is radius-mass relation for self-gravitating 
objects supported by degenerate neutrino pressure, assuming three neutrino
types of mass 2 ev.  The dashed line is the same when the neutrino mass is
taken to be 1 ev.  The dotted line separates the Newtonian and MOND regimes
and the points are relevant to the dark matter cores of the sample clusters.
The closed points are those clusters with significant discrepancies as before.}
\vfill
\end{figure}

I have calculated the structure of a self-gravitating system 
supported by the degenerate pressure of neutrinos (a neutrino star).
In the non-relativistic case there is a pressure-density relation of
the form $p_\nu = K{\rho_\nu}^{5/3}$ where $K = 5.5\times 10^{32}(m_\nu/
2\, ev)^{-{8\over 3}}$,
and density is in g/cm$^3$.  Inserting this relation in eq.\ 1, I determine
the run of density in a system with a given central velocity dispersion.
Such objects (n=1.5 polytropes) have nearly constant density out to a finite
radius where the density drops to zero.  As for white dwarfs, there is a 
mass-radius relation.  This is shown by the solid curve ($m_\nu = 2$ ev)
and the dashed curve ($m_\nu = 1$ ev) 
in Fig.\ 9 where I have assumed that the central
velocity dispersion of the neutrinos is equal to the gas velocity dispersion.
The solid points show the mass and radius of the dark matter cores for
those clusters with significant discrepancies
($M_d/M_g>1.5$) and the open points those with small discrepancies.  The dotted
line separates the MOND and Newtonian regime where we see that the mass-radius
relation has different forms:  $r\propto M^{1/12}$ in the MOND regime and 
$r\propto M^{-1/3}$ in the Newtonian regime.  It is of interest
that for $M_\nu = 2$ ev the range of radii and masses correspond in magnitude
to the dark cores required in clusters.  Moreover, we see that those clusters
with large discrepancies lie generally above the mass-radius relation as would
be expected for neutrino clusters with density less than or equal to that given
by eq.\ 13.  Further, not even neutrinos of 2 ev apparently could 
contribute to the mass budget of those objects with low discrepancies.

That neutrinos in the range of one to two ev make up the missing mass in 
clusters is a provocative 
possibility, and one which is within reach of experimental verification.  The 
accelerator limits on the electron mass can soon be pushed to within a few 
tenths of an electron volt;  if there is no positive detection, then active 
neutrinos cannot supply the dark mass of clusters.

\section {Conclusions}
 
What one may call, generically, "the dark matter problem" first became evident
with radial velocity studies of rich clusters of galaxies (Zwicky 1933).
The discrepancy between the visible and Newtonian
dynamical mass, quantified then in
terms of mass-to-light ratio, was more than a factor of 100.  With the
advent of X-ray observatories and the detection of hot 
gas in clusters, this discrepancy between dynamical and detectable
mass was reduced to a factor of 10.  Modified Newtonian dynamics reduces the
discrepancy further, to a factor of 2 to 3, but it is clear that a
discrepancy remains which cannot be explained by detected gaseous or luminous
mass.  It is also apparent that while the missing mass is primarily in the
inner regions of clusters, it does extend beyond the core radius as defined by
the gas density distribution.

How serious is this remaining mass discrepancy for MOND?  Formally speaking,
it  does not constitute a falsification.  If the dynamical mass
predicted by MOND were generally less than the detected mass in stars and gas, 
then it would be a definite falsification, but this is not the case.  More mass
can always be found (as in the case of the hot gas), but it is difficult
to make observed mass disappear.

One might reasonably argue that the implied presence of undetected mass runs 
against the spirit of MOND, which, in the view of most people, was suggested 
primarily as a replacement for dark matter.  But more generally, MOND should
not be viewed simply as an alternative to dark matter;  the
systematic appearance of the mass discrepancy in astronomical systems with
low internal accelerations is an indication that Newtonian dynamics or gravity
may break down in this limit.  MOND primarily addresses this issue: is physics
in the low-acceleration regime Newtonian? The success of MOND in explaining the
scaling properties and observed rotation curves of galaxies suggests that it
may not be.  MOND does not rest upon the principle that there is no undetected
or dark matter.  Indeed, comparing the density of luminous matter to the
baryonic content of the universe implied by considerations of primordial
nucleosynthesis, one can only conclude that there is, as yet, undetected
baryonic matter, probably in the form of diffuse gas in the intergalactic
medium.  Moreover, it is virtually certain that particle dark matter exists in
the form of neutrinos; only its contribution to the total mass density of the
Universe is unclear.

MOND would be incompatible with the wide-spread existence of
dark matter which clusters on the scale of galaxies-- cold dark matter. 
But MOND is not inconsistent with hot dark matter such as 2 ev
neutrinos, which can only aggregate on the scale of clusters of galaxies--
indeed, I have presented evidence that this may be the case.  
Neutrinos, as particle dark matter candidates, are unquestionably 
well-motivated, both from a theoretical point-of-view (they definitely exist) 
and from an experimental point-of-view (they have mass).  No conjectured CDM
particle shares these advantages.  While I do not wish to state that the
dark matter in clusters is definitely in the form of 2 ev neutrinos (there
is more than enough remaining baryonic matter to make up the missing mass), there
are indications that point this way.  The largest discrepancies are found
in the clusters with the largest core radii as would be the case with neutrinos.
The indicated densities of dark matter are comparable to the maximum possible
density of 2 ev neutrinos.  The radius-mass relationship for self-gravitating
degenerate neutrino objects forms an envelope for those clusters with large
discrepancies.

The fact remains that there exists an algorithm, MOND, which allows galaxy
rotation curves to be predicted in detail from the observed distribution of
matter, and it is for these systems that the kinematic
observations are most precise.
This fact challenges the current CDM paradigm, and demands explanation if
dark matter lies behind the discrepancy.  The factor two remaining discrepancy
in clusters is less challenging for MOND, particularly given that 
MOND makes no claims about the full material content of the Universe.

I am grateful to Hans B\"ohringer for useful discussions and for a copy
of the thesis of T.H. Reiprich.  I thank Moti Milgrom for helpful comments
on the manuscript.

\end{document}